\documentclass[aps,twocolumn,nofootinbib,showpacs,amsmath,superscriptaddress]{revtex4-1}
\usepackage{graphicx}
\usepackage{epstopdf}
\usepackage{bm}
\usepackage{epsfig}
\usepackage{graphics}
\usepackage{xspace}

\def\OMIT#1{}

\newcommand{\nn}{\nonumber}

\newcommand{\bn}{{\bar n}}
\newcommand{\bea}{\begin{eqnarray}}
\newcommand{\eea}{\end{eqnarray}}

\newcommand{\gsim}{\mathrel{\rlap{\lower4pt\hbox{\hskip1pt$\sim$}}\raise1pt\hbox{$>$}}}

\newcommand{\be}{\begin{equation}}
\newcommand{\ee}{\end{equation}}

\begin{document}
\title{N-Jettiness as a Probe of Nuclear Dynamics}

\author{Zhong-Bo Kang}
\affiliation{RIKEN BNL Research Center, 
                   Brookhaven National Laboratory, 
                   Upton, NY 11973}

\author{Sonny Mantry}
\affiliation{High Energy Division, 
                  Argonne National Laboratory, 
                  Argonne, IL 60439}
\affiliation{Department of Physics and Astronomy, 
                   Northwestern University,
                   Evanston, IL 60208}

\author{Jian-Wei Qiu}
\affiliation{Physics Department, 
                   Brookhaven National Laboratory, 
                   Upton, NY 11973}
\affiliation{C.N. Yang Institute for Theoretical Physics, 
                   Stony Brook University, 
                   Stony Brook, NY 11794}

\begin{abstract}
We propose the use of $N$-jettiness ($\tau_N$), a global event shape, as a probe of nuclear dynamics in lepton-nucleus collisions. It characterizes the amount of soft radiation between the jet and nuclear beam directions. We give the factorization for the 1-jettiness ($\tau_1$) distribution for the production of a single hard jet ($J$) in lepton-nucleus collisions: $\ell+A(P) \to J(P_{J})+X$. Each nuclear target gives rise to a unique pattern of radiation, affected by nuclear dynamics, that can be quantified by the $\tau_1$-distribution.   Up to power corrections, the $\tau_1$-distribution allows for a direct measurement of the nuclear PDFs.  Additional nuclear-dependent effects will be dominated through power corrections of size  $\sim Q_s^2(A)/(\tau_1P_{JT})$ where $Q_s(A)$ is a dynamical nuclear scale and $P_{JT}$ is the transverse momentum of the jet.   Such nuclear-dependent effects can be studied through a dedicated program to measure $\tau_1$-distributions for a range of nuclei and kinematics. We give numerical results for the 1-jettiness distribution for the simplest case of a proton target at next-to-leading-log accuracy.

\end{abstract}


\maketitle

Event shape distributions have played a vital role in advancing our understanding of various dynamical aspects of QCD.
 A new global event shape for exclusive $N$-jet cross-sections, called $N$-jettiness ($\tau_{N}$)~\cite{Stewart:2010tn,Jouttenus:2011wh}, was recently introduced to veto additional jets in an inclusive manner. The definition of $\tau_{N}$ is given by
\bea
\label{tauN}
\tau_N &=& \sum_{k} \text{min}_{i} \Big \{ \frac{2 q_i\cdot p_k }{Q_i}\Big \},
\eea
where the set  $\{q_i\}$ denote reference four-vectors along the beam and the $N$ jet directions. The sum
over $k$ runs over all the hadronic final state particles of momentum $p_k$, and the $Q_i$ are arbitrary normalization constants on the order of the hard scale in the process. Different choices of the set $Q_i$ correspond to different definitions of $\tau_N$ with different properties~\cite{Jouttenus:2011wh}. The region of small $\tau_N$ corresponds to events with very little radiation between the $N$ jets and the beam directions, with the limit $\tau_N \to 0$ corresponding to $N$ infinitely narrow and well separated jets.

For processes with nuclei in the initial state, N-jettiness can provide a quantitative way to characterize the unique pattern of final state radiation thereby probing nuclear medium effects. N-jettiness can thus be used as a new  diagnostic tool for studying various aspects of nuclear dynamics. Jet tomography is known to be an important tool to diagnose properties of the quark-gluon plasma (QGP) produced in relativistic heavy ion collisions \cite{jet-quenching}.   The interactions of an energetic jet moving through a dense medium change the momentum spectrum of  jets  (jet-quenching) leading to the spectacular phenomenon of leading particle or jet suppression observed in relativistic heavy ion collisions at both RHIC and the LHC \cite{data-jet-quenching}. The same interaction also induces additional radiation to alter the radiation pattern between the beam and jet directions and to change the overall jet shape.  Instead of varying jet shape parameters to observe such effects \cite{vitev-etal}, one can alternatively use  N-jettiness as a way to quantify the radiation pattern, 
while keeping the jet definition unchanged.  The combination of jet-quenching and N-jettiness measurements can provide a comprehensive jet tomography to diagnose the properties of a dense medium, such as QGP or the medium of ordinary nuclei. Similarly, nuclear dynamics in other processes such as nuclear deep inelastic scattering (DIS) can be studied through N-jettiness distributions.

In order to compute $\tau_N$~\cite{Stewart:2010tn, Jouttenus:2011wh}, any standard infrared-safe jet algorithm can be employed to obtain $N$ jets  and the resulting jet momenta are used as the reference jet vectors in the set $\{q_i\}$ in Eq.~(\ref{tauN}).   The only information needed from the jet algorithm for the calculation of $\tau_N$ are these reference jet momenta. In the limit $\tau_N\to 0$, where the jets are well separated, different jet algorithms will give rise to the same set of reference momenta. The jet algorithm dependence is thus power suppressed in the region of small $\tau_N$. The reference momenta $\{q_i\}$ can also be obtained directly from a minimization procedure for $\tau_N$ without the use of a jet algorithm~\cite{Thaler:2011gf}. 

In the N-jettiness formalism, all final state particles are grouped either with one of the beam directions or one of the jet directions through the minimization requirement in Eq.~(\ref{tauN}). The $i$-th jet is then defined to be the set of particles grouped in the $q_i$-th direction. Note that the contribution of particles collinear with one of the reference vectors $q_i$ or particles with soft momenta is suppressed due to the dot products $\{q_i\cdot p_k\}$ in Eq.~(\ref{tauN}). The largest contributions arise from particles with sizable momenta in a direction distinct from the set $\{q_i\}$. This allows one to isolate events with N narrow jets with only soft radiation between the jets by restricting to a region of small $\tau_N$. One of the advantages of  N-jettiness  over other event shapes such as Thrust is that it allows one to be exclusive in the number of final state jets while simultaneously being insensitive to the contribution from forward beam remnants, which are experimentally difficult to control. The forward beam remnant contribution to $\tau_N$ is suppressed, since these particles are mostly collinear with the beam reference momenta in the set $\{q_i\}$.  Furthermore, $\tau_N$ quantifies the shape of radiation between the N jets as opposed  to variables like Thrust which do not distinguish the various jet and beam regions. For more details on the N-jettiness formalism we refer the reader to Refs.~\cite{Stewart:2010tn,Jouttenus:2011wh} and references within. 

As a first step, in this paper, we consider a specific application of N-jettiness: single jet ($N=1$) production in lepton-nucleus collisions. Such an analysis can be generalized for multiple jets ($N>1$) in a straightforward manner. Inclusive production of  a single jet with a high transverse momentum ($P_{JT}$)  and rapidity ($y$) in lepton-nucleus collisions, $\ell+A(P) \to J(P_{J})+X$, is a well-defined observable and can be systematically calculated in the QCD collinear factorization formalism  \cite{Kang:2011jw}.  We set up a factorization formalism, based on the Soft Collinear Effective Theory (SCET) \cite{SCET},  that in addition gives the 1-jettiness distribution for such processes. The value of $\tau_1$ reflects the amount of radiation between the jet and nuclear beam directions. By studying the $\tau_1$ distribution for a wide range of nuclei, one can systematically probe the effect of the nuclear environment on the observed pattern of radiation. For larger nuclei, the $\tau_1$-distribution is expected to be broader with the peak position shifted toward larger values of $\tau_1$. This corresponds to the enhanced hadronic activity between the jet and beam directions  due to  nuclear-medium effects in larger nuclei. Such a program can be carried out at the proposed Electron-Ion Collider (EIC) where a wide range of nuclear targets are planned 
\cite{Boer:2011fh}. In particular, a measurement of the ratio of the cross-sections
\bea
\label{ratio}
\frac{d\sigma(A, \tau_1,P_{J_T},y)}{d\sigma(A=1,\tau_1,P_{J_T},y)},
\eea
between a larger nuclear target (A) and the nucleon target (A=1) can isolate A-dependent nuclear medium effects in the three dimensional configuration space ($\tau_1,P_{J_T},y$), allowing for systematic studies of dynamical nuclear effects. Many of the nuclear-independent theoretical uncertainties are likely to cancel in the ratio in Eq.(\ref{ratio}).

Without requiring the detection of the scattered lepton, the jet transverse momentum 
$P_{JT}$ is the only hard scale in this process.  The tree-level partonic scattering is given by lepton-quark scattering through a virtual photon exchange $\ell+q\to \ell'+q$. The $\tau_1$ distribution is just proportional to $\delta(\tau_1)$, corresponding to the infinitely narrow quark jet, and takes the form
\bea
\label{PM}
\frac{d^3\sigma^{(0)}}{dy dP_{JT} d\tau_1}=\sigma_0\, \delta(\tau_1) \sum_q e_q^2\, f_{q/A}(x) ,
\eea
where $\sigma_0\equiv \frac{d\hat{\sigma}}{dy\>dP_{JT}}$ is just the tree-level partonic cross-section differential in the jet rapidity and transverse momentum. The sum over $q$ runs over the initial quarks and antiquarks, $e_q$ is the corresponding fractional charge, $f_{q/A}$ is the corresponding nuclear 
parton distribution function (PDF) \cite{nPDF},  $\sigma_0$ and initial parton momentum fraction $x$ are given by
\bea
\label{normx}
\sigma_0=\frac{4\pi\alpha_{em}}{Q^3e^y}\left[\frac{\hat s^2+\hat u^2}{\hat t^2}\right], \quad
x=\frac{e^{y}P_{JT}/Q }{1-e^{-y}P_{JT}/Q},
\eea
respectively, with $\hat s$, $\hat t$, and $\hat u$ the usual partonic Mandelstam variables, and $Q=\sqrt{s}$ denotes the lepton-nucleon center-of-mass (CM) energy. Non-zero values of $\tau_1$ will occur only beyond tree level in perturbation theory. For $\tau_1\ll P_{J_T}$, fixed-order perturbation theory breaks down due to the appearance of large logarithms of the form of $\alpha_s^n\ln^{2n} \left(P_{JT}/\tau_1\right)$. A systematic resummation of such logarithms can be performed~\cite{Stewart:2010tn} in the SCET framework. Furthermore, non-perturbative effects will modify the tree-level result in Eq.(\ref{PM}) and these effects will also be incorporated.

We work in the lepton-nucleus  CM frame. At leading order in the SCET power counting, the factorization formula for the $\tau_1$-distribution takes the schematic form
 \bea
\label{fac}
\frac{d^3\sigma}{dy dP_{JT} d\tau_1} \sim H\otimes B_A\otimes J\otimes {\cal S},
\eea
where $\otimes$ represents a convolution, $H$ is the hard function encoding physics of the hard scattering, $B_A$ is the nuclear beam function~\cite{Stewart:2009yx} encoding the dynamics of the initial state nucleus and the collinear beam radiation, $J$ is the jet function for the dynamics of collinear modes along the jet direction, and ${\cal S}$ is a soft function for soft radiation throughout the event. We refer the reader to Ref.~\cite{Stewart:2009yx} for more details on the beam function and how it differs from the jet function. The invariant mass of the nuclear beam jet and the final state jet is characterized by the scale $\sqrt{\tau_1 P_{JT}}$ and the soft radiation has momentum of size $\tau_1$. The relevant scales in the problem are then given by
\bea
\label{scales}
P_{JT} \gg \sqrt{\tau_1 P_{JT}} \gg \tau_1,
\eea
so that $H$, $B_A$, $J$, and ${\cal S}$ naturally live at the scales $\mu_H\sim P_{JT}$, $\mu_J\sim \sqrt{\tau_1 P_{JT}}$, and $\mu_S \sim \tau_1$ respectively. For perturbative values of $\mu_J$, the beam function $B_A$ can be matched \cite{Stewart:2009yx} onto the usual nuclear 
PDFs $f_A$ as
\bea
\label{beam}
B_A \sim {\cal I} \otimes f_A,
\eea
where the perturbative coefficient ${\cal I}$ is independent of the nucleus. 

According to Eqs.~(\ref{fac}) and (\ref{beam}), for $\tau_1\gg \Lambda_{\rm QCD}$, the $\tau_1$-distribution is completely determined in terms of the nuclear PDFs and the perturbatively calculable functions  $H$, ${\cal I}$, and ${\cal S}$. For $\tau_1\sim \Lambda_{\rm QCD}$, the universal soft function ${\cal S}$ becomes non-perturbative. The soft function ${\cal S}$ is independent of the nuclear target and can thus  be extracted from the data on lepton-nucleon scattering and used as an input for scattering off heavier nuclei. Furthemore, for kinematics where $\mu_J \sim \sqrt{\tau_1 P_{J_T}}\sim \Lambda_{QCD}$, the beam, jet and soft functions all become non-perturbative. Such effects can be important for $\tau_1 \sim \Lambda_{QCD}^2/P_{J_T}\ll \Lambda_{QCD}$.

The leading power formalism is modified through power corrections. The power corrections are characterized through ratios of the scales $\mu_H,\mu_J,\mu_S,$ and $Q_s(A)$. 
$Q_s(A)$ is a dynamical scale which characterizes the size of power corrections from multiple scattering in a nuclear medium and is a function of the nuclear atomic number. For the simplest case of the nucleon, $Q_s^2(A=1) \sim \Lambda_{QCD}^2$. For larger nuclei $Q_s^2(A) \sim A^\alpha \Lambda_{QCD}^2$ where $\alpha$ denotes the power law dependence with atomic number $A$.  If there is no color transfer between nucleons,  $\alpha$ is expected~\cite{Luo:1994np} to be $\sim 1/3$ corresponding to the length of multiple scatterings in the nucleus.  When the scales $\mu_H$ or $\mu_J$ are of the same order as $Q_s(A)$, the power corrections cannot be ignored and a new approach such as the Color Glass Condensate approach \cite{Gelis:2010nm} is needed. $Q_s(A)$ is also sometimes referred as the saturation scale \cite{Gelis:2010nm}.  

Power corrections to the hard function correspond to multiple hard scatterings at the scale $\mu_H \sim P_{J_T}$ \cite{Qiu:1990xy}. Such effects  generate power suppressed operators which give rise to higher twist nuclear beam functions which are then matched onto higher twist nuclear matrix elements \cite{Luo:1994np}.  Thus, these effects can also give rise to nuclear dependent effects but will be suppressed by the hard scale $\mu_H\sim P_{J_T}$. The dominant nuclear-dependent power corrections arise from corrections to the OPE in Eq.(\ref{beam}) and scale as $Q_s^2(A)/\mu_J^2$. These effects correspond to multiple scatterings along the nuclear beam at the jet scale with typical size
\bea
\frac{Q_s^2}{\mu_J^2} \sim \frac{\Lambda_{\rm QCD}^2\, A^{\alpha}}{\tau_1 P_{JT}}\, ,
\eea
and can be  enhanced by the large number of gluons~\cite{Gelis:2010nm} in a large nucleus.
Precision measurements of the $\tau_1$-distribution for different values of $A$ and $P_{J_T}$ can  allow for an independent extraction of $\alpha$ and  test the expectation~\cite{Luo:1994np,Boer:2011fh} of $\alpha\sim 1/3$. Power corrections to the jet and soft functions will be universal and independent of the nuclear target. However, the convolutions of these universal power corrections with the nuclear beam functions can give rise to differences in the $\tau_1$-distribution for different nuclei.

 By comparing data with the leading-power factorization formalism in Eqs.~(\ref{fac}) and (\ref{beam}), any measured non-trivial nuclear modification to the $\tau_1$-distribution is  strong evidence of coherent multiple scattering, and can be of interest in the study of multi-parton correlation, energy loss in cold nuclear matter, etc.

We now give some details on the SCET framework and the resulting factorization formula. In the lepton-nucleus CM frame, the collinear degrees of freedom along the nuclear beam have momenta proportional to the light-cone vector $n_A^\mu=(1,0,0,1)$. We define a conjugate light-cone vector 
$\bn_A^\mu$  so that $n_A^2=\bn_A^2=0, n_A\cdot \bn_A = 2$, and $\vec{n}_A=-\vec{\bn}_A$. 
Similarly,  the light-cone four-momentum vectors $n_J^\mu$ and $\bn_J^\mu$ associated with the jet direction so that $ n_J^2=\bn_J^2=0$, $n_J\cdot \bn_J = 2$,  and $\vec{n}_J=-\vec{\bn}_J$, with the jet direction along $\vec{n}_J$.

The collinear modes along the nuclear beam and jet directions have momentum scalings 
\bea
&&(n_A \cdot p,\bn_A \cdot p,  p^{\perp_A} ) \sim P_{JT} ( \lambda^2,1,  \lambda),\nn \\
&&(n_J \cdot p,\bn_J \cdot p,  p^{\perp_J}) \sim P_{JT} (\lambda^2,1,  \lambda),
\eea
respectively where $\lambda\sim \sqrt{\tau_1/P_{JT}}$ and $p^{\perp_A}$ and $p^{\perp_J}$are momentum components perpendicular to the beam and jet axis respectively. The soft radiation has momentum scaling 
\bea
(n_A \cdot p, \bn_A \cdot p,  p^{\perp_A} )\sim P_{JT} (\lambda^2, \lambda^2, \lambda^2).
\eea
The factorization formula for the differential cross-section derived using SCET is given by
\bea
\label{factorization}
\frac{d^3\sigma}{dy dP_{JT} d\tau_1} &=&\sigma_0 \sum_{q,i} e_q^2 \int_0^1 dx\int_x^1 \frac{dz}{z} \int ds_J \int dt_a  
\nn \\
&&\times H(x Q P_{J_T}e^{-y}, \mu_J; \mu_H) \nn \\
&& \times\delta \big [ x- \frac{e^y P_{J_T}}{Q-e^{-y}P_{J_T}}\big ]  J^q(s_J, \mu_J)
\nn \\
&&\times {\cal I}^{qi}\left(\frac{x}{z}, t_a, \mu_J\right) f_{i/A}(z,\mu_J) 
 \\
&&\times {\cal S}\left(\tau_1 - \frac{t_a}{Q_a}-\frac{s_J}{Q_J}, \mu_J;\mu_S\right),\nn
\eea
which is the detailed version of the schematic formulas in Eqs.~(\ref{fac}) and (\ref{beam}) and $Q_a=x Q, Q_J=\bar{n}_J\cdot P_{J}$. In deriving the above formula we also used the reference vectors $q_A^\mu=x P_A^\mu$, where $P_A$ is the initial state nucleus momentum and $q_J=(P_{J_T}\cosh y, \vec{P}_{J_T},P_{J_T}\sinh y)$ which is just the momentum of a massless jet with transverse momentum and rapidity $P_{J_T}$ and $y$ respectively. The last two arguments of the hard function $H$  denote that renormalization group (RG) evolution between the scales $\mu_H$ and $\mu_J$ is included. Similarly, the soft function is evolved between the scales $\mu_S$ and $\mu_J$. The jet function $J^q$, the beam function to PDF matching coefficient ${\cal I}^{qi}$, and the nuclear PDF $f_{i/A}$  are all evaluated at the common jet scale $\mu_J$. The tree-level values for various functions above are $H=1$, $J^q=\delta(s_J)$, ${\cal I}^{qi}=\delta_{qi} \delta(t_a)\delta(1-x/z)$
and ${\cal S}=\delta(\tau_1 - \frac{t_a}{Q_a}-\frac{s_J}{Q_J})$. For field theoretic definitions of the beam, jet, and soft functions we refer the reader to \cite{Jouttenus:2011wh} and references within.
Large logarithms in the ratio of the scales $\mu_H \sim P_{JT}$, $\mu_J\sim \sqrt{P_{JT}\tau_1}$, and $\mu_S \sim \tau_1$ are resummed using the RG equations in SCET. 

In phenomenological studies, theoretical uncertainties to Eq.(\ref{factorization}) will arise from a truncation of the perturbative series in the hard, jet, and soft functions, the power corrections discussed earlier, and from non-perturbative soft function effects when $\tau_1\lesssim \Lambda_{QCD}$. In addition, Eq.(\ref{factorization}) will be affected by the standard PDF uncertainties. In the following, we make numerical predictions for the $\tau_1$-distributions for the simplest case of a proton target. We estimate the perturbative uncertainties via a standard scale variation. We study the effects of non-perturbative soft radiation by modeling the soft function in the region $\tau_1 \lesssim \Lambda_{QCD}$ and varying the model parameters as described below. Numerical studies of power corrections are left for future work. Any deviations between data and the prediction of Eq.~(\ref{factorization}), after including the remaining uncertainties, will be a measure of the size of power corrections. These power corrections can be studied as a function of $(\tau_1,y,P_{JT})$. In the ratio in Eq.~(\ref{ratio}), the uncertainties from nuclear-independent effects will largely cancel out so that a deviation from unity probes  nuclear-dependent effects.

\begin{figure}
\includegraphics[height=2in,width=3in]{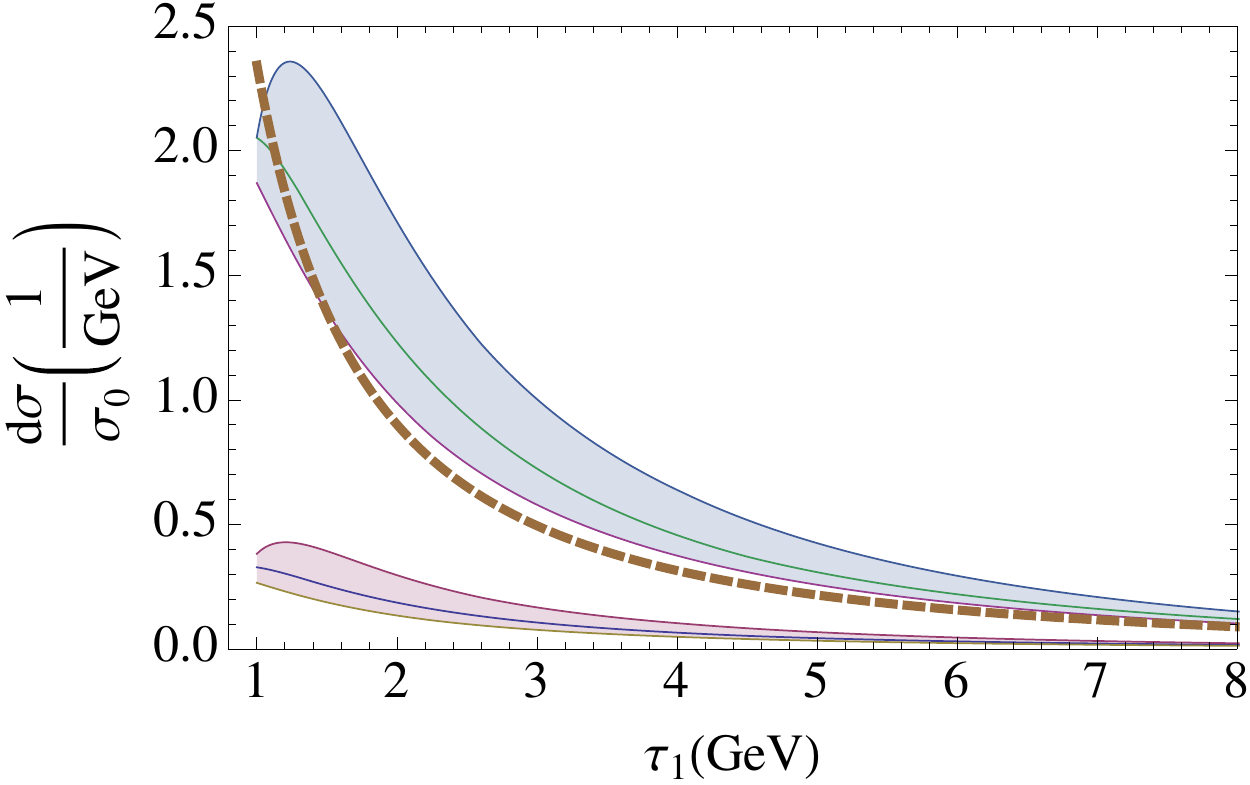}
\caption{$d\sigma / \sigma_0\equiv \frac{1}{\sigma_0}\frac{d^3\sigma}{dy dP_{JT} d\tau_1}  $ as a function $\tau_1$ for a proton target at NLL accuracy. The bottom  and top bands correspond to $\sqrt{s}=90 \>\text{GeV}, P_{JT}=20 \>\text{GeV}, y=0$ (Stage I EIC) and    $\sqrt{s}=300\>\text{GeV}, P_{JT}=20 \>\text{GeV},y=0$ (HERA) respectively.  The dashed curve shows the singular part of the NLO cross-section for HERA kinematics. The scale choices for  $\mu_H,\mu_J,\mu_S$ are explained in the text. }
\label{heraeic}
\end{figure}

 Fig.~\ref{heraeic} shows the $\tau_1$-distribution at Next-to-Leading-Log (NLL) accuracy for typical Stage-I EIC and HERA kinematics. We follow the conventions in Table 1 in Ref.~\cite{Berger:2010xi} for counting logs, set $n_f=5$, and use CTEQ6L PDFs~\cite{Nadolsky:2008zw}. The shaded bands correspond to NLL scale variation by choosing $\mu_H = r P_{JT}, \mu_J= r \sqrt{P_{JT} \tau_1}, \mu_S = r\> \tau_1$ for the range $r=\{1/2,2\}$. The curve in the middle of each band corresponds to $r=1$. 
We also show (dashed curve) the singular part of the NLO cross-section for HERA kinematics. The effect of resummation is to tame the singular behavior of the fixed order cross-section in the $\tau_1\to 0$ limit. The $\tau_1$-distribution is cutoff at 1 GeV so that the soft function is still perturbative.

In the region $\tau_1 \sim \Lambda_{\rm QCD}$, the soft function ${\cal S}(\tau_1,\mu)$ becomes non-perturbative and is modeled as
\bea
{\cal S}(\tau_1,\mu)= \int dk_a dk_J \delta (\tau_1 - k_a - k_J) {\cal S}(k_a,k_J,\mu), 
\eea
where ${\cal S}(k_a,k_J,\mu)$ is the generalized hemisphere soft function~\cite{Jouttenus:2011wh}. Non-perturbative physics in this hemisphere soft function is encoded by the convolution of the partonic soft function with a model function $S_{\rm mod}$~\cite{Ligeti:2008ac,Hoang:2007vb,Fleming:2007xt} as 
\bea
\label{softmodel}
 {\cal S}(k_a,k_J,\mu) &=& \int dk_a' dk_j'  \>S_{\rm mod} (k_a',k_J')
 \nn\\
 &&\times{\cal S}_{\rm part.}(k_a-k_a',k_J-k_J',\mu), 
\eea
where ${\cal S}_{\rm part.}$ corresponds to the perturbative soft function.
$S_{\rm mod}(k_a',k_b')$ is chosen to peak near $k_{a,b}'\sim \Lambda_{\rm QCD}$ so that for $\tau_1\gg \Lambda_{\rm QCD}$ the soft function reduces to the perturbative result ${\cal S}_{\rm part.}$. For NLL resummation, the $\tau_1$ distribution is sensitive to the non-perturbative physics only through the combination 
\bea
F_{\rm mod}(u) =\int_{-u}^u \frac{d\zeta}{2} \>S_{\rm mod}\left(\frac{u+\zeta}{2},\frac{u-\zeta}{2}\right).
\eea 
This result is derived  by expanding Eq.~(\ref{factorization}) to the NLL  level and using  $u=k_a'+k_b', \zeta=k_a'-k_b'$ in Eq.~(\ref{softmodel}).

\begin{figure}
\includegraphics[height=2.3in,width=3.2in]{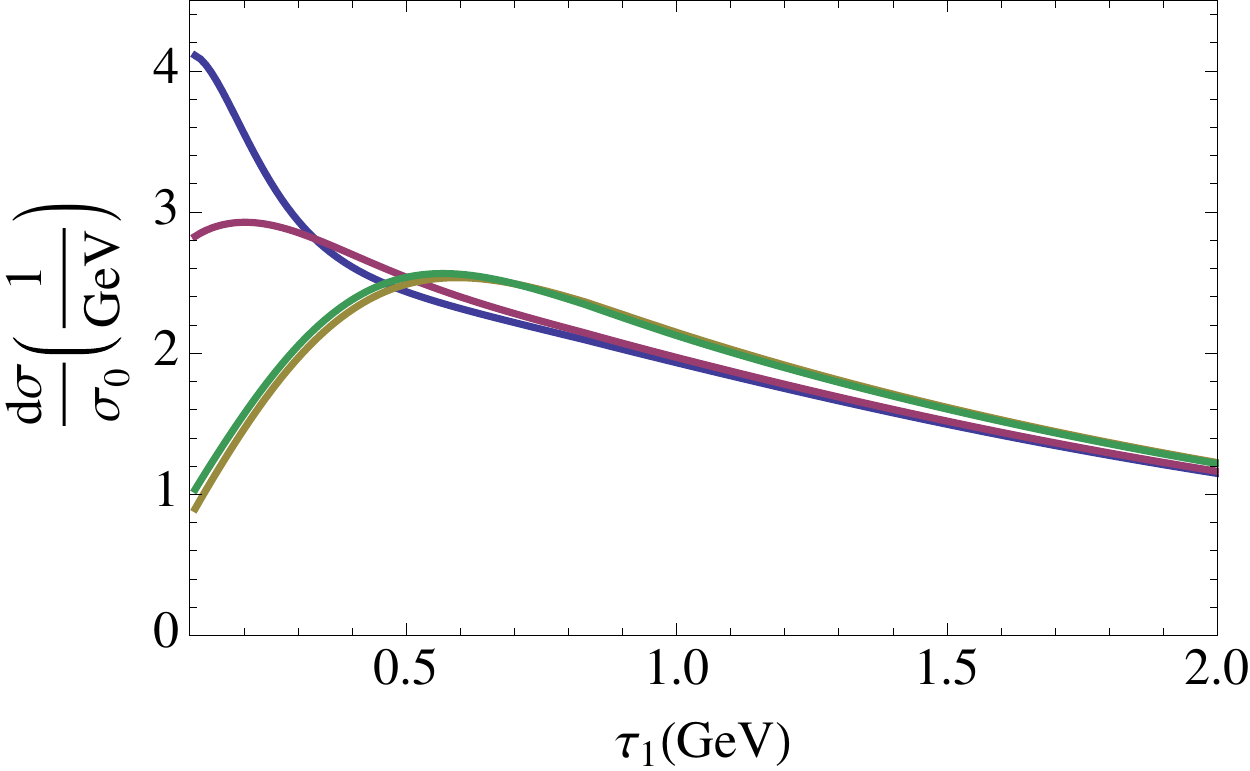}
\caption{$d\sigma / \sigma_0\equiv \frac{1}{\sigma_0}\frac{d^3\sigma}{dy dP_{JT} d\tau_1}  $ as a function $\tau_1$ for a proton target at NLL accuracy including non-perturbative $\tau_1$ values with $\sqrt{s}=300$ GeV,  $P_{JT}= 20$ GeV, and $y=0$. The different curves with peak positions from left to right correspond to $(a,b,\Lambda)=(2.0, -0.2, 0.2),  (1.2, -0.1, 0.3), (2.2, -0.4, 0.5),\\(1.8, -0.05, 0.4)$ in Eq.~(\ref{fmod}) respectively. }
\label{NPmodels}
\end{figure}

The function $F_{\rm mod}$ is parameterized as 
\bea
\label{fmod}
F_{\rm mod}(u) = \frac{N(a,b,\Lambda)}{\Lambda}\Big (\frac{u}{\Lambda}\Big )^{a-1}{\rm Exp}\left[-\frac{(u-b)^2}{\Lambda^2}\right],
\eea
where the normalization factor $N(a,b,\Lambda)$ is chosen to satisfy $\int_0^\infty du F_{\rm mod}(u)=1$, equivalent to the condition $\int dk_a \int dk_J {\cal S}_{\rm mod}(k_a,k_J) =1$. In Fig.~\ref{NPmodels}, we plot the 1-jettiness cross section as a function of $\tau_1$ for different parameter choices of  $a$, $b$, and $\Lambda$ as explained in the caption. We made the scale choices $\mu_H=P_{J_T},\mu_J=\sqrt{\tau_1 P_{J_T}}$, and $\mu_S=\tau_1\sqrt{1+(\tau_{1}^{min}/\tau_1)^2}$ with $\tau_1^{min}=1$ GeV. The different models exhibit different behavior for $\tau_1\sim \Lambda_{\rm QCD}$ but converge to the perturbative result for larger values of $\tau_1$ as required. The universality of the soft function allows its extraction from lepton-nucleon collisions to be then used in scattering off larger nuclei.

A similar analysis can be performed for other nuclear targets. We leave further studies of nuclear phenomenology, higher order resummation, and power corrections for future work. 
  
 We thank Xiaohui Liu and Frank Petriello for useful discussions. This work was supported in part by by the U.S. Department of Energy under Contract No.~DE-AC02-98CH10886 (ZK and JQ) and by the U.S. National Science Foundation under grant NSF-PHY-0705682 (SM).

\end{document}